\documentclass{mem}
\usepackage{natbib}
\usepackage{txfonts}
\usepackage{balance}
\usepackage{graphicx}
\usepackage{flushend}
\usepackage{siunitx}
\usepackage[breaklinks,pdftex]{hyperref}
\idline{75}{282}
\begin{document}
\def\teff{$T\rm_{eff }$}
\def\kms{$\mathrm {km s}^{-1}$}
\newcommand{\daoafrac}{\frac{\Delta\alpha}{\alpha}}
\newcommand{\daoaline}{\Delta\alpha/\alpha}
\newcommand{\cmps}[1]{\SI{#1}{\centi\meter\per\second}}
\newcommand{\mps}[1]{\SI{#1}{\meter\per\second}}
\newcommand{\kmps}[1]{\SI{#1}{\kilo\meter\per\second}}

\newcommand{\comm}[1]{{\color{magenta}{#1}}}

\newcommand{\MgI}{Mg{\sc \,i}~}
\newcommand{\MgII}{Mg{\sc \,ii}~}
\newcommand{\MnII}{Mn{\sc \,ii}~}
\newcommand{\FeII}{Fe{\sc \,ii}~}
\newcommand{\FeI}{Fe{\sc \,i}~}
\newcommand{\NiII}{Ni{\sc \,ii}~}
\newcommand{\ZnII}{Zn{\sc \,ii}~}
\newcommand{\aivpfit}{\texttt{AI-VPFIT}}
\title{
Methods for quasar absorption system measurements of the fine structure constant in the 2020s and beyond
}

% ALTERNATIVE TITLES
% Developments in measurements of the fine structure constant at high redshift
% Methods for measurements of the fine structure constant in 2020s and beyond
% Methods for astronomical measurements of the fine structure constant in 2020s and beyond
% Measurements of the fine structure constant in the 2020s and beyond

   \subtitle{}

\author{
D. \,Milakovi{\'c}\inst{1,2,3} 
\and 
C.C. \,Lee\inst{4}
\and 
P. \,Molaro\inst{2,1}
\and
J.K. \,Webb\inst{4}
          }

\institute{
Institute for Fundamental Physics of the Universe, Via Beirut 2, I-34532 Trieste, Italy
\and
INAF -- Osservatorio Astronomico di Trieste, Via Tiepolo 11,
I-34131 Trieste, Italy
\and
INFN, Sezione di Trieste, Via Valerio 2, I-34127 Trieste, Italy
\and
Clare Hall, University of Cambridge, Herschel Rd, Cambridge CB3 9AL, United Kingdom
\\
\email{dinko@milakovic.net, dmilakov@sissa.it}
}

\authorrunning{Milakovi{\'c} et al.}

\titlerunning{Methods for $\alpha$ measurements in quasars at high redshift}

\date{Received: Day Month Year; Accepted: Day Month Year}

\abstract{
This article reviews the two major recent developments that significantly improved cosmological measurements of fundamental constants derived from high resolution quasar spectroscopy. The first one is the deployment of astronomical Laser Frequency Combs on high resolution spectrographs and the second one is the development of spectral analysis tools based on Artificial Intelligence methods. The former all but eliminated the previously dominant source of instrumental uncertainty whereas the latter established optimal methods for measuring the fine structure constant ($\alpha$) in quasar absorption systems. The methods can be used on data collected by the new ESPRESSO spectrograph and the future ANDES spectrograph on the Extremely Large Telescope to produce unbiased $\daoaline$ measurements with unprecedented precision. 
}
\maketitle{}

\section{Introduction}

The fine structure constant, $\alpha\equiv{e^2}/{\hbar c}$ is a dimensionless parameter of the Standard Model of Particle Physics whose value determines the strength of the electromagnetic interactions. As such, it is considered one of the fundamental constants of nature. Variations of fundamental constants with time are possible, and even required, in higher-dimensional theories (e.g.\ string and Kaluza-Klein) and in theories in which a scalar field $\phi$ provides late-time acceleration of the universal expansion rate (e.g.\ quintessence and Bekenstein type models). Detecting any such variations would disprove the Weak Equivalence Principle, a postulate of General Relativity. \cite{Uzan2011} and \cite{Martins2017} give an overview of theoretical models associated with fundamental constant variations.  \par

Any deviation in $\alpha$ would cause a small, systematic shift in the energy levels of atoms and thus their wavelengths. This shift can equivalently be expressed as a small spectral velocity shift, $\Delta v$:
\begin{equation}
    \frac{\Delta v}{c} = - \frac{\Delta \alpha}{\alpha} \frac{2q}{\omega}. 
\end{equation}
Above, $\omega$ is the wavenumber of the transition, $q$ quantifies its sensitivity to $\alpha$ variation, $c$ is the speed of light, and 
\begin{equation}
    \frac{\Delta \alpha}{\alpha} = \frac{\alpha_{obs} - \alpha_{lab}}{\alpha_{lab}}, 
\end{equation}
where $\alpha_{obs}$ and $\alpha_{lab}$ refer to the values in the observed astronomical object and the laboratory, respectively. 

$q$-coefficients have been determined through many-body quantum mechanical calculations \citep[e.g.\ ][]{Dzuba1999}. Measuring relative velocity shifts between a set of multiplets, covering a wide range of $q$-coefficient values, is the basis of the Many Multiplet (MM) method \citep{Webb1999,Dzuba1999}. To illustrate the required spectroscopic wavelength accuracy for such measurements, a 1 part-per-million (ppm) change in $\daoaline$ produces a relative velocity shift of $\approx \mps{20}$ between \FeII$\lambda2344$ and \MgII$\lambda2796$ transitions, both commonly observed in quasar absorption systems. 
\par

The MM method was applied to quasar absorption system \citep{Webb1999,Webb2001}, white dwarf \citep{Berengut2013,Bainbridge2017,Hu2021}, and stellar spectra \citep{Hees2020,Murphy2022_solartwins}. The application of the MM method to 293 quasar absorption systems (observed by VLT/UVES and Keck/HIRES) found evidence for a spatial dipole with an amplitude of $\approx 10$\,ppm and a $4.2\sigma$ statistical significance \citep{King2012}. 

Later studies revealed the existence of short- and long-range wavelength-scale distortions in both UVES and HIRES \citep{Rahmani2013,Whitmore2015}. Left uncorrected, such distortions spoil $\daoaline$ measurements. While incorporating the knowledge on the distortions and the associated uncertainty is possible \citep{Dumont2017}, it is clear that significant advancements to wavelength calibration of astronomical spectrographs is necessary to advance the field. \par

Quasar absorption systems commonly consist of clustered gas clouds which appear as a complex structure of blended Voigt profiles in the spectrum. It is possible to produce statistically acceptable models (in terms of $\chi^2$ statistics) of the system with different -- and discrepant -- $\daoaline$ measurements \citep{Wilczynska2015}. \par

Information criteria (see Section~\ref{sec:IC}) can discriminate between competing models, but modelling complex absorption systems is often time consuming for humans. Additionally, a human may unconsciously bias the results towards some preferred value. Unbiased and automated spectral fitting tools are necessary to explore these issues and ensure robust measurements. \par

In this paper, we review recent advancements in instrumentation delivered by the Laser Frequency Comb technology (Sec.~\ref{sec:lfc}), by the ESPRESSO spectrograph (Sec.~\ref{sec:espresso}), and data analysis tools based on Artificial Intelligence, \aivpfit~(Sec.~\ref{sec:analysis}). We then present some lessons learned by applying \aivpfit~to simulated and real quasar spectra in Sec.~\ref{sec:insights}, with conclusions in Sec.~\ref{sec:discussion}. \par 

\section{Laser Frequency Combs}\label{sec:lfc}
The Laser Frequency Comb \citep[LFC, ][]{Udem2002,Haensch2006} in an ideal wavelength calibrator for high-resolution spectrographs. It produces a set of unresolved emission lines equidistant in frequency space. The frequency of an $n^{\rm th}$ line, $f_n$, is given by:
\begin{equation}
    f_n = f_0 + n\times f_r,
\end{equation}
The ``offset'' and ``repetition'' frequencies ($f_0$ and $f_r$, respectively) are stabilised to an external radio frequency standard with an accuracy of $\Delta f/f\sim10^{-12}$ or, equivalently, \SI{2}{\milli\meter\per\second} \citep{Probst2020}. 

Astronomical LFCs  have $f_r\approx \SI{20}{\giga\hertz}$, approximately a factor of three above their instrumental resolution element width. Imaging the LFC spectrum provides thousands of densely spaced lines on the detector. The measured line positions, together with their known frequencies and thus wavelengths, can be used to wavelength calibrate the spectrograph and remove distortions to better than $ \Delta\lambda/\lambda\simeq 10^{-8}$ or, equivalently, \mps{3} \citep{Milakovic2020}. Whilst there are other potential sources of uncertainty, reducing the wavelength calibration errors to this level can reduce its contribution to the overall error budget on $\daoaline$ to $\leq0.1$ppm.

\subsection{Lessons from HARPS}\label{sec:lfc:harps}

Following a development period \citep{Wilken2010,Wilken2012}, in 2015 HARPS \citep{Mayor2003} became the first high resolution spectrograph to be equipped with an LFC for regular operations \citep{Probst2014}. The LFC performance was characterised in a series of tests described here. 

The use of both static and dynamic mode scrambling methods in optical fibres carrying LFC light was found to be crucial for obtaining a uniform and stable line spread function and therefore for obtaining wavelength calibration stability at the photon noise limit \citep{Probst2020}. When fitting Gaussians to LFC lines, it was found that line centres shift by $\approx\mps{1}$ between consecutive exposures when dynamic scrambling is not used \citep{Probst2020}. Furthermore, the measured Gaussian line centres were found to depend on the flux level, due to charge transfer inefficiency (CTI) of the detector, with shifts as large as \mps{3} for sufficiently low fluxes \citep{Milakovic2020,Zhao2021}. Using static and dynamic scrambling and applying a simple CTI correction, the HARPS LFC demonstrated a \mps{3} wavelength calibration accuracy (the difference between the wavelength calibration polynomial and the known LFC line wavelength, averaged over $\sim 10\, {\rm k}$ lines) in a single exposure and $<\cmps{1}$ wavelength calibration stability in a series of 194 exposures taken over period of several hours \citep{Milakovic2020}.

A disturbing result emerged from the simultaneous use of two LFCs on HARPS \citep{Probst2020,Milakovic2020}. The two LFCs shared the same basic design but had different $f_0$ and $f_r$ and were operated independently. The comparison of the wavelength calibrations from the two LFCs revealed an unexpected $60\pm \cmps{1}$ discrepancy in their zero-points. Applying a simple CTI correction decreased the discrepancy to $ \mps{45}$ \citep[c.f. the thin and thick black and red lines in figure 15 of ][]{Milakovic2020}. Further improvements could be achieved using a more accurate CTI model and possibly by employing the empirical instrumental profile instead of its Gaussian approximation (see Sec.~\ref{sec:espresso:ip} regarding the latter).

Although the tests discussed above have not identified any systematic effects associated with the use LFC itself, the origin of these shifts is not yet fully understood. A tunable LFC (i.e. one for which $f_0$ or $f_r$ can be precisely controlled) may be useful in this respect. Such a system can be used to change the LFC line positions at will and, in this way, scan the detector in sub-pixel steps. In addition to producing more detailed CTI models and reconstructing the instrumental profile on very small wavelength scales, this could also be used to quantify the impact of intra-pixel sensitivity variations on line centre measurements.

\section{ESPRESSO}\label{sec:espresso}
ESPRESSO is a high resolution ($R=\lambda/\delta\lambda=145$k, where $\delta\lambda$ is the full-width at half-maximum of the resolution element) spectrograph designed for measurements at the photon noise limit, with $\alpha$ measurements being one of its main scientific goals \citep{Pepe2021}. Wavelength calibration errors were reduced to a negligible level in ESPRESSO using an LFC \citep{Schmidt2021,Murphy2022}. In addition, ESPRESSO is also equipped with a thorium-argon (ThAr) lamp which is observed simultaneously with a Fabry-P{\'e}rot etalon (FP) for high precision calibration. 

\begin{figure}
\resizebox{\hsize}{!}{\includegraphics[clip=true]{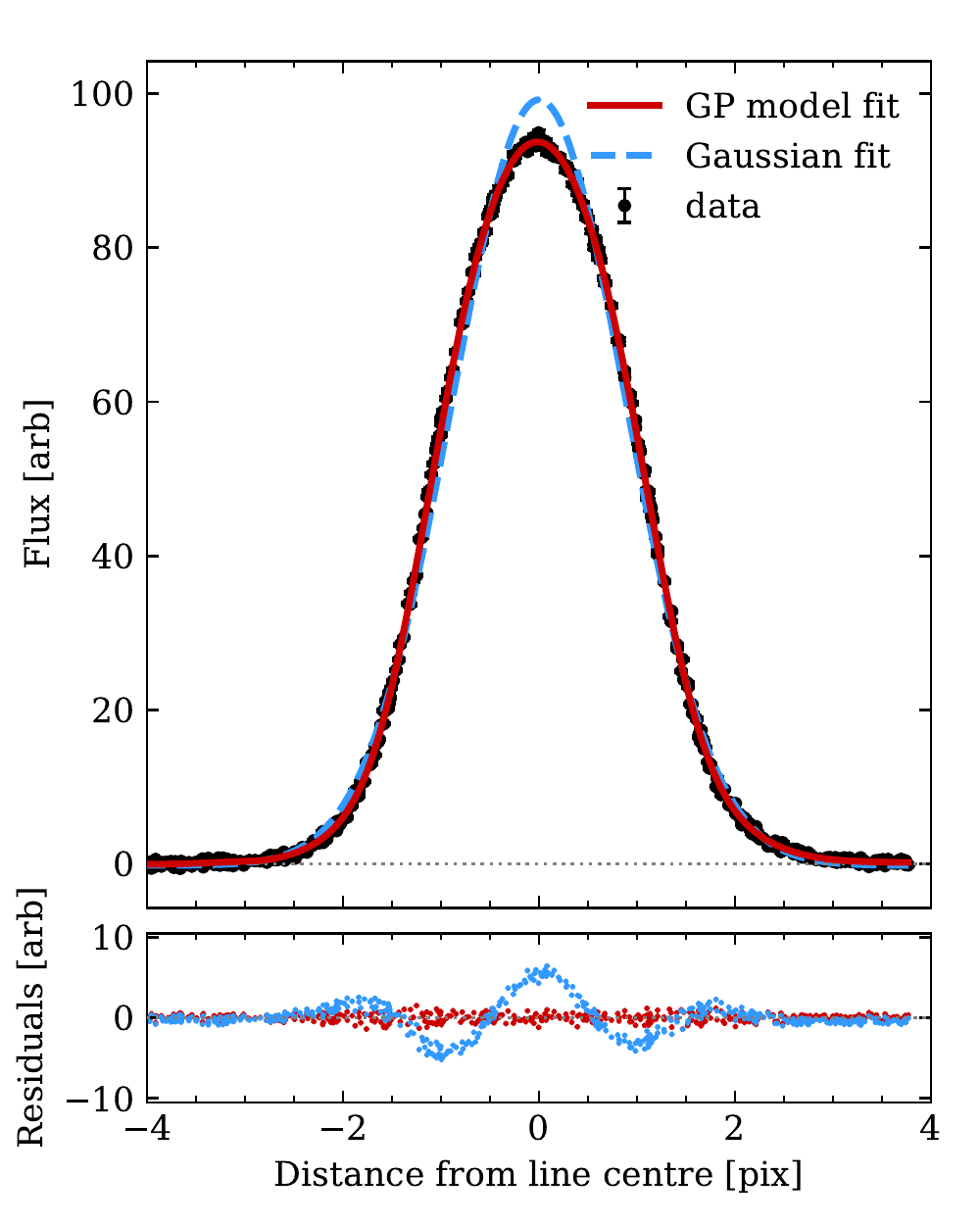}}
\caption{
\footnotesize
{\it Top panel}: LFC lines in the region $500\pm\SI{0.25}{\nano\meter}$, observed in one slice of ESPRESSO, were flux normalised and stacked on their Gaussian profile centres (black points, $N=520$). The reconstructed, non-parametric, non-Gaussian IP is shown as a solid red line. The reconstruction was performed using Gaussian process (GP) regression on the black data points (Milakovi{\'c} et al., in prep.). A Gaussian IP fitted to the same data is shown as a dashed blue line for comparison. The difference in the modes of the two fitted profiles is 0.01\,pix, corresponding to $\approx \mps{4}$. {\it Bottom panel}: Residuals to the two IPs fitted above. Residuals to the non-parametric IP are shown in red. Asymmetry of the empirical IP is seen as a wavy pattern in the residuals to the Gaussian IP approximation, shown in blue. }
 
\label{fig:lsf}
\end{figure}

\begin{figure}
\resizebox{\hsize}{!}{\includegraphics[clip=true]{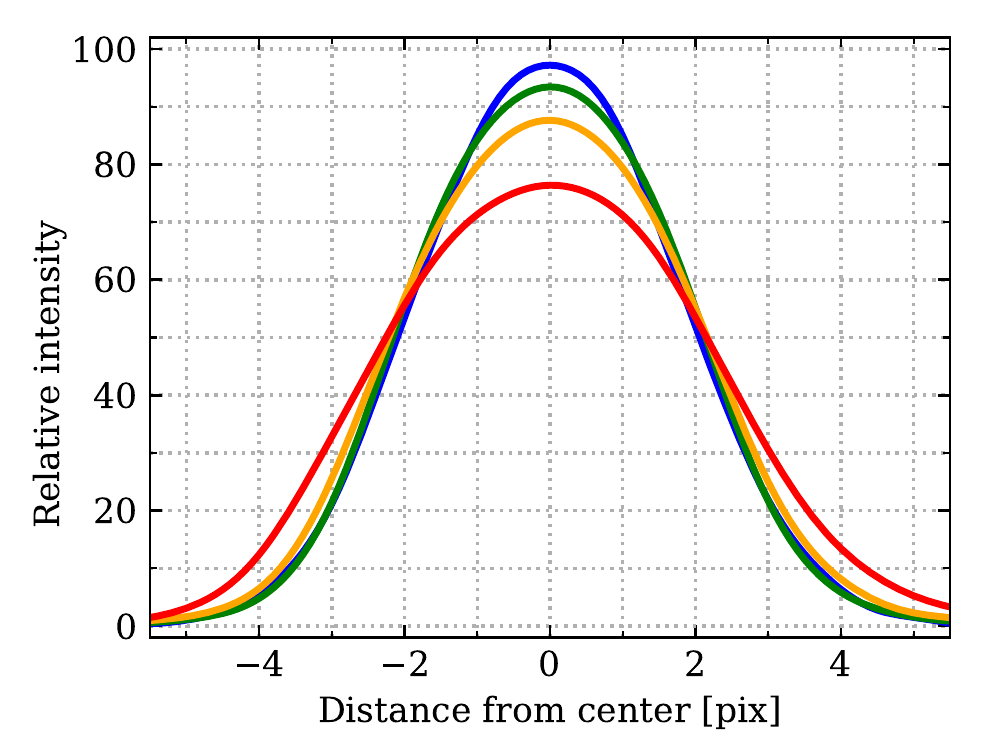}}
\caption{
\footnotesize
The coloured lines show the empirical IP of ESPRESSO in a single slice of the optical order 122 (central wavelength \SI{500}{\nano\meter}). Different colours correspond to different sections of the order, changing from blue to red with increasing wavelength. 
}
\label{fig:lsf_order}
\end{figure}

\subsection{Instrumental profile reconstruction}\label{sec:espresso:ip}
Another significant advantage of the LFC is the narrow intrinsic width of its lines, $\approx \SI{1}{\mega\hertz}$ (c.f. full-width at half maximum of the instrument resolution element $\approx\SI{5}{\giga\hertz}$), before passing through the instrument optics. As such, images of LFC lines on the detector directly trace the instrumental profile (IP). Reconstruction of ESPRESSO's IP (Fig.~\ref{fig:lsf}) was done in a non-parametric way, i.e.\ using Gaussian process regression \citep{Rasmussen2006}. Preliminary results revealed that the IP is asymmetric in pixel space and that the asymmetry varies as a function of position on the detector (Fig.~\ref{fig:lsf_order}). Fitting the LFC lines using the empirical IP results in differences in the line mode as large as \mps{80} with respect to when a Gaussian IP is assumed (Milakovi{\'c} et al., in prep.). For comparison, the desired accuracy in measuring line centroids in $\daoaline$ studies is \mps{5}. 

Knowledge of the empirical IP can be used to improve wavelength calibration accuracy, better understand instrumental effects, and in spectral modelling. For example, some of the problems with ESPRESSO wavelength calibration (discussed in Section~\ref{sec:espresso:lfc_limitations}) may be resolved by better knowledge of the line centres. Importantly, they may also explain the observed difference in the zero-points of two independent LFCs on HARPS (Sec.~\ref{sec:lfc:harps}). Lastly, Voigt profile fitting tools such as \texttt{VPFIT} \citep{Carswell2014} and \aivpfit~(see Sec.~\ref{sec:analysis}) already allow for the use of an empirical IP so, as soon as the IP models are validated, they will be used in future analyses.

\subsection{Limitations of the ESPRESSO LFC}\label{sec:espresso:lfc_limitations}
Since its installation in 2018, ESPRESSO's LFC demonstrated a series of issues preventing its use for regular instrument calibrations and monitoring. Hardware issues caused the LFC's spectral coverage and line intensities to vary on timescales of minutes, preventing a reliable wavelength calibration for the instrument. Additionally, a detailed study of the ESPRESSO's LFC identified other unexpected problems preventing its performance at theoretical limits \citep{Schmidt2021}. 

Firstly, the ThAr+FP and LFC wavelength calibrations differ up to \mps{30} within a single echelle order, higher than expected considering the 2-\mps{4} uncertainties on the positions of LFC and ThAr lines \citep[cf.\ figure 13 of ][]{Schmidt2021}. This issue probably points to inaccuracies of ThAr atlases but has not been investigated further. Secondly, the residuals to the LFC (but also ThAr+FP) wavelength calibration solution show peculiar properties. The histogram of the residuals is double-peaked and the standard deviation of the residuals around the best fit wavelength calibration solution show excess scatter: approximately \mps{6} (68\% c.l.) instead of the expected \mps{2} \citep[cf.\ figure 10 of ][]{Schmidt2021}. Thirdly, residuals of consecutive LFC lines are correlated on short pixel-scales \citep[cf.\ figure 9 of][]{Schmidt2021}. The mentioned effects, while unexpected, are unlikely to spoil $\daoaline$ measurements. 

Finally, technological limitations constrain ESPRESSO LFC's spectral range to the region between \SI{490} and \SI{720}{\nano\meter}, approximately 43\% of the wavelength range of ESPRESSO. Important ultraviolet transitions, e.g.\ \FeII\,$\lambda1608$ and $\lambda1611$, are therefore not covered at $z<2$. Additionally, the lack of LFC coverage limits the wavelength range in which IP can be determined. This is an additional motivation for the development of a new generation of astronomical LFCs \citep{Obrzud2019,Obrzud2021,Schmidt2021bluves}. Instrument servicing in May 2022 and in October 2022 aimed to resolve the technical issues. Preliminary results show a broader wavelength coverage and more stable flux levels.

\begin{figure}
\resizebox{\hsize}{!}{\includegraphics[clip=true]{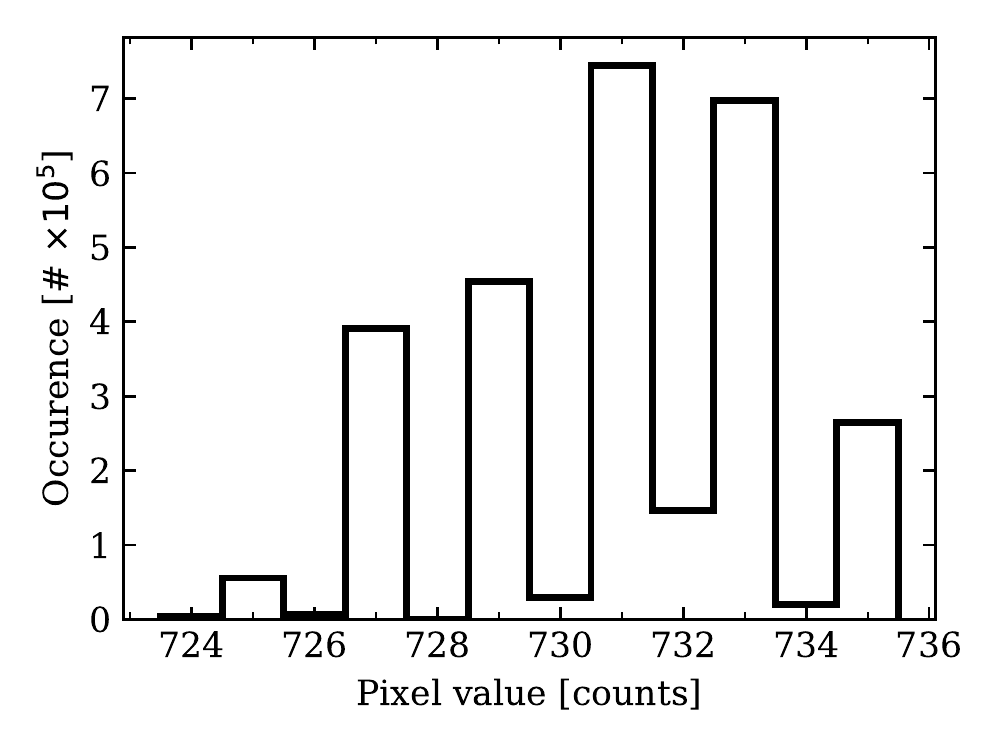}}
\caption{
\footnotesize
A histogram of raw pixel values in a single read-out region of ESPRESSO blue detector's bias frame taken on 3rd November 2018 at 10.11 UTC. Approximately 93\% (7)\% of the values are odd (even) as opposed to the expected 50/50\% occurrence. The issue affects all read-out regions and appears in all ESPRESSO raw frames. See text in Sec.~\ref{sec:feb} for more details.
}
\label{fig:binary}
\end{figure}

\subsection{A digitization issue in ESPRESSO before May 2022}\label{sec:feb}
All exposures taken between ESPRESSO's commissioning in November 2018 and 19 May 2022 were affected by a digitization issue caused by a malfunction in the Next Generation Controller boards \citep{Baade2009}. The problem was first reported in \cite{Murphy2022_messenger} and resolved quickly after being detected in late 2021, as explained shortly.
The malfunction caused individual segments of the detector, each corresponding exactly to a single read-out region, to preferentially output odd or even numbers. The issue was tracked to a small variability in the reference voltage used in the ADC (Antonio Manescau, priv.\ comm.) In the worst affected segments, such as the one shown in Fig.~\ref{fig:binary}, 7\% (93\%) of pixels had even (odd) values. 

Raw detector frames thus contained pixels for which the read-out value is off by one with respect to their true values. The affected pixels seem to vary randomly from one exposure to the next. The amplitude of the effect does not depend on the recorded flux levels in the exposure. The impact of this effect on science measurements derived from quasar spectra has not yet been assessed. The boards were since replaced and observations taken after 19 May 2022 do not suffer from this problem.

\section{Spectral modelling using Artificial Intelligence}\label{sec:analysis}
Recent years saw the development of novel spectral modelling tools designed to produce objective, robust and unbiased $\daoaline$ measurements: \texttt{GVPFIT} \citep{Bainbridge2017} and \aivpfit~ \citep{Lee2021aivpfit}. 

\aivpfit~combines non-linear least squares minimisation, genetic algorithms, Monte Carlo (MC), and Artificial Intelligence (AI) methods to automatise the process of absorption system modelling. Its genetic algorithm relies on an information criterion to choose the best-fit model in each generation and to decide when to stop the modelling process. The use of high performance computing increased the efficiency by a factor of $>100$ over a human modeller: a model of an arbitrarily complex absorption system can be produced in several hours or days using supercomputers\footnote{A human would require several weeks or months for completing the same task.}. This allowed a series of checks on the assumptions made in previous studies, revealing biases and sources of systematic uncertainty, as discussed in this section. Importantly, it also identified ways to avoid them in future measurements.

\subsection{Model building based on information theory}\label{sec:IC}
The optimal model should strike a balance between simplicity (i.e.\ the number of model parameters) and parameter bias and variance, as discussed in \cite{Bozdogan1987, Webb2021}. An information criterion (IC) penalises the introduction of unnecessary model parameters by adding a penalty term ($\mathcal{P}$) to the $\chi^2$ statistic, $\textrm{IC}=\chi^2 + \mathcal{P}$. When comparing several competing models, each of which gives $\chi^2/\nu\simeq1$ (where $\nu$ is the number of degrees of freedom), the preferred model is the one that minimizes the IC. This way, the IC allows for objective model selection based on information theory \citep[see, e.g., ][]{Liddle2007}. 

\aivpfit~allows the user a choice between three IC: 
\begin{enumerate}
    \item the corrected Akaike Information Criterion (AICc), $\mathcal{P}={2kN}/({N-k-1})$ \citep{Akaike1974,Hurvich1989}.
    \item the Bayesian Information Criterion (BIC), $\mathcal{P}=k\ln(N)$ \citep{Schwarz1978}; 
    \item the Spectroscopic Information Criterion (SpIC), for which the exact $\mathcal{P}$ formula is in \cite{Webb2021}. 
\end{enumerate}
Above, $k$ and $N$ are the total number of free model parameters and the number of data points, respectively. 

In a typical quasar absorption system modelled by \aivpfit, AICc tends to overfit the data and BIC tends to underfit it, while SpIC strikes a balance between the two \citep[c.f.\ fig 2 in ][]{Webb2021}. The greatest benefit of SpIC is that its formulation considers the ``local'' impact of model parameters by taking into account the number of pixels they affect and the strength of the spectral feature being fitted. This is crucial in order to avoid systematic effects in $\daoaline$ measurements, as discussed in Section~\ref{sec:insights}.

\subsection{The importance of the Monte Carlo approach}\label{sec:mc}

The introduction of Monte Carlo approaches in conjunction with \aivpfit~offer several vital advantages. However, first it is helpful to clarify what exactly the Monte Carlo aspects are. Inherent to the \aivpfit~itself is the randomised way in which trial components are added to the absorption complex, during model construction. There is no ``intelligent'' placement of components, which may at first sight seem inconsistent with an interactive human approach. Interactively, a human may be more likely to target the strongest visible component, or if not, perhaps the ``least blended'' component. Clearly, throwing down components randomly does not necessarily achieve this human selection procedure, although equally clearly, {\it some} of the trial positions in a Monte Carlo sense {\it will} replicate the interactive approach. 

But there is second Monte Carlo aspect to \aivpfit~usage; one can form a large number of models of any given absorption system, each time using a different random number seed for trial component placement. In this way, a large number of independently constructed models has the capacity to incorporate all possible human modellers. 

Together, these two Monte Carlo procedures not only permit a completely unbiased approach to model construction, but also enable an exploration of parameter space so that multiple minima in $\chi^2$ space can be mapped out. At the time of writing this paper, relatively few absorption systems have been modelled using \aivpfit. Therefore, how common multiple minima are is not yet known. Nevertheless, on the basis of what has been done, it seems that the phenomenon is likely to be common \citep{Lee2021nonunique,Webb2022}. 

The discussion above naturally raises the question as to what {\it causes} multiple minima in $\chi^2$ space in the context of $\daoaline$ measurements. For example, would some completely different statistical procedure (e.g. machine learning) identify the same local minima as does non-linear least squares? The answer to this is not known. How sensitive is the shape of $\chi^2$ space to spectral S/N and resolution? Again, this has not yet been quantified. Nevertheless, it is clear that individual error bars \citep[e.g. see the blue points in figure 6 of ][]{Webb2022} will grow with decreasing S/N and lower resolution, such that local minima will be less pronounced or indistinguishable. 

\section{Insights from AI-VPFIT}\label{sec:insights}

Several thousand \aivpfit~models were constructed for both simulated and real quasar spectra and the impact of particular choices for the model building process were investigated in a series of papers by \cite{Lee2021aivpfit,Lee2021nonunique,Webb2021,Webb2022}. These include assumptions on the gas broadening mechanism, the choice of the IC, the decision to hold the $\daoaline$ fixed throughout the model building process. The following section outlines the main results of the four publications mentioned in this paragraph. 
\subsection{On the line broadening mechanism}\label{sec:insights:broadening}
Assumptions on absorbing gas motion proved to be crucial in obtaining robust $\daoaline$ measurements. Thermal and turbulent motions in the absorbing gas cloud impact on the width of its observed Voigt profile ($b$-parameter) such that the total, compound, $b$-parameter is given by:
\begin{equation}
    b^2_{\mathrm{tot}} = b^2_{\mathrm{turb}} + b^2_{\mathrm{therm}},
\end{equation}
where $b_{\mathrm{turb}}$ arises due to turbulence and $b^2_{\mathrm{therm}}={2k_BT}/{m}$ depends on gas temperature $T$ and atomic mass $m$, with $k_B$ the Boltzmann's constant. 

Simulations showed that superfluous components are required to obtain a good fit to the data when compound broadening is not permitted, and that the extra components often emulate compound broadening. Furthermore, because of degeneracies introduced by extra components' parameters, the statistical uncertainty\footnote{Calculated from the diagonal of the covariance matrix at the best fit solution.} on $\daoaline$ increases. Allowing for compound broadening was the most reliable way to retrieve the input parameters. The converse was also true: assuming an incorrect line broadening mechanism may generate false local minima and hence produce a spurious result  \citep{Lee2021aivpfit, Webb2022}.

\subsection{On model non-uniqueness}\label{sec:insights:nonuniqueness}

Model non-uniqueness was further explored by \cite{Lee2021nonunique} using HARPS observations of the quasar HE0515$-$4414 \citep[$R=115\mathrm{k}$, data described in ][]{Milakovic2021} and UVES observations of the quasar Q0528$-$2505 \citep[$R=70\mathrm{k}$, data described in ][]{Murphy2019}). Several hundred models were made for each of the two quasars using different combinations of the line broadening mechanism (turbulent, thermal, or compound) and the IC (AICc or SpIC) and their resulting $\daoaline$ measurements compared.

The main results of that analysis can be seen in their figure 1. For both quasars, non-uniqueness was present for certain combinations of the broadening mechanism and IC. When non-uniqueness was present, the scatter in $\daoaline$ was at least comparable to the average statistical uncertainty (see, e.g., panels (a) and (d) of the aforementioned figure). Non-uniqueness was the most severe when AICc and turbulent broadening were used (c.f.\ panels (a) and (g) of the same figure). Yet, when SpIC and compound broadening were used, HE0515$-$4414 showed no evidence for non-uniqueness (c.f.\ panels (a) and (e)). The benefit of SpIC and compound broadening was less evident for Q0528$-$2505 (c.f.\ panels (g) through (l)), possibly because of the lower resolution of the UVES observations. 

The combination of using SpIC and compound broadening significantly decreased the probability of ending up in a local minimum, therefore lowering the systematic uncertainty associated with non-uniqueness in the analysed HARPS observations of HE0515$-$4414. The opposite was true for all other combinations of IC and gas broadening \citep[as seen from the large scatter of points in the corresponding panels of figure 1 in ][]{Lee2021nonunique}.

\subsection{On bias}\label{sec:insights:bias}
One of the conclusions from \aivpfit~is that modelling procedures matter. Assumptions on gas broadening and the choice of IC are directly (but partially) related to the probability of ending up in a local minimum and hence the level of bias. This raises an important question regarding avoid bias in future measurements. %If they are biased, what should be changed to avoid it?

Prior to the {\sc ai-vpfit} studies that have revealed that substantial bias may actually be introduced, one approach attempting to avoid biasing $\daoaline$ measurements has been to fix $\daoaline=0$ during model building. Only once the model has been obtained in this way was $\daoaline$ included as a model parameter. The assumptions here are that the procedure just described does not cause the model to get stuck in a local minimum, i.e. that an unbiased value of $\daoaline$ is always recovered.
This approach was followed in almost all published analyses \citep[e.g.\ ][]{Webb1999,Webb2001,King2012,Kotus2017,Murphy2022}. 

Going a step further, some authors applied additional methods to their data in order to avoid biasing results \citep[e.g.\ ][]{Evans2014,Kotus2017,Murphy2017, Murphy2022}. Weak long-range and intra-order distortions were introduced into the wavelength-scales of individual quasar exposures before their combination, effectively offsetting $\daoaline$ from its true value by $\lesssim 5\,\mathrm{ppm}$ \citep{Murphy2022}. The offset magnitude is random and unknown outside of the software used to introduce the distortions \citep[\texttt{UVES\_popler}, ][]{Murphy2018UVESpopler}. The modelling is done on the distorted spectrum in the way described above. The only exception happens at the very end, when the final optimisation of model parameters is done on the original, undistorted data instead.

However, \cite{Webb2022} have shown that the combination of fixing $\daoaline =0$ and applying distortions is particularly vulnerable to bias on $\daoaline$ when turbulent broadening is used to simultaneously model transitions arising in atoms of different masses. The reason for that is twofold. Firstly, if (for example) the absorption components being modelled are in reality thermally broadened but are fitted using turbulently broadened profiles, the final model will be forced to contain more components to achieve a good fit to the data. The presence of superfluous absorbing components creates more degeneracy i.e. the individual parameters are less well constrained. Secondly, if $\daoaline$ is fixed to be zero during model construction, the parameters describing those superfluous absorbing components can more readily adjust to give a good fit to the data. The net result is that the model obtained in this way is likely to end up in a false local minimum. It has previously been assumed that the final step of releasing $\daoaline$ as a free parameter can ``release'' the model from its false local minimum and iterate to the true value of $\daoaline$, whatever that may be. We now know this is wrong and in fact, once {\it in} a local minimum, it is likely that the model stays there, ending up with a spurious result.

The bias caused by fixing $\daoaline =0$ and applying distortions was investigated using a simulated ESPRESSO spectrum of HE0515$-$4414 \citep{Webb2022}. The simulated spectrum was created with $\daoaline=8.08\,\textrm{ppm}$ (a value previously recovered from the real data). One hundred \aivpfit~models were created, each with $\daoaline$ set to zero, for each of the four combinations of turbulent/compound broadening and AICc/SpIC. $\daoaline$ was only allowed to vary (together with other model parameters) in the final optimisation step. 

Considering compound broadening only, and referring to figure 6 in \cite{Webb2022} (i.e. the red points), we can see good consistency between the individual measurements. There seems to be no bias (the measurements reflect the input value of $\daoaline$) and there are no multiple minima. Conversely, considering only turbulent broadening (i.e. the blue points), we see enhanced scatter as well as clear multiple minima. 
In fact, a $\daoaline$ value consistent with the input value was recovered 100\% of the time only when compound broadening and SpIC were used, with compound broadening plus AICc almost as good. A measurement consistent with the input $\daoaline$ was recovered only $\approx10$\% of the time when turbulent broadening and AICc was used. Instead, a measurement consistent with zero was found $\approx50\%$ of the time. Given that the average statistical uncertainty in those models was $\langle\sigma_\textrm{stat}\rangle\approx2.5\,\rm{ppm}$, a no-variation result is $\sim 3\sigma$ away from the simulated input.

\section{Discussion \& conclusions}
\label{sec:discussion}
This article reviews a series of publications that have advanced measurements of the fine structure constant over the last several years. The most significant of them are the deployment of astronomical Laser Frequency Combs for wavelength calibration and the development of \aivpfit~for spectral analysis. The former led to more accurate wavelength calibration of the quasar spectra from which the measurements are derived and the latter led to significant improvements in measurement methods. 

The application of LFC wavelength calibration methods on HARPS and ESPRESSO removed the previously dominant source of instrumental uncertainty in $\daoaline$ measurements (Sec.~\ref{sec:espresso}).
The development of \aivpfit~(Sec.~\ref{sec:analysis}) has been equally groundbreaking. Not only did it increase the speed at which measurements can be made, but it identified methods to avoid non-uniqueness and reduce measurement bias (Sec.~\ref{sec:insights}). 

ESPRESSO is expected to observe dozens of quasars during its operational lifetime, producing many more high quality, high S/N spectra suitable for varying constant studies. The developments described here will be useful for the analysis of that data, but also of the data collected using the ANDES spectrograph \citep{Marconi2022}, to be installed on the Extremely Large Telescope in the future.

The conclusions, summarised as enumerated points below, are derived mostly from published work referenced in the text. The references to the original work from which conclusions numbered 1--7 are derived can be found in the sections therein stated. Conclusions numbered 8, 9, and 10 are derived jointly from the four publications referenced at the beginning of Section~\ref{sec:insights}. For this reason, claims mentioned in conclusions 8--10, but originally presented elsewhere, are given their exact references. 

The main conclusions are:

\begin{enumerate}
    \item  Instruments calibrated using LFCs are free from wavelength-scale distortions at the level of $\Delta\lambda/\lambda\simeq 10^{-8}$ (Sec.~\ref{sec:lfc}). This translates into a maximum systematic uncertainty on $\daoaline$ of approximately $<0.1\,\textrm{ppm}$ due to wavelength calibration errors. Even in the absence of an LFC, distortions in the ThAr+FP wavelength calibration on ESPRESSO are only marginally worse (Sec.~\ref{sec:espresso}). 

    \item LFCs are also excellent tools for characterising instrumental effects (e.g.\ CTI, Sec.~\ref{sec:lfc:harps}) and reconstructing the empirical IP (Sec.~\ref{sec:espresso:ip}). Tunable LFCs may increase their utility as tools for instrument characterisation.
    
    \item Empirical models of the ESPRESSO IP will further improve wavelength calibration accuracy and hence reduce associated systematics (Sec.~\ref{sec:espresso:ip}). When provided to software such as \aivpfit~, the IP models will lead to more accurate models of absorption systems. 
    
    \item Developments in LFC technology are expected to address outstanding issues like LFC reliability and wavelength coverage (Sec.~\ref{sec:espresso:lfc_limitations}). 
    
    \item The zero-points of two independent LFCs observed on HARPS differ by $\cmps{45}$ (Sec.~\ref{sec:lfc:harps}). This has significant implications on science cases requiring $\sim\cmps{1}$ instrument stability over periods $>1$\,year. Future instruments which will rely on LFC to track instrument drifts, such as NIRPS \citep{Bouchy2017} and ANDES \citep{Marconi2022}, could cross-check instrument stability by using redundant calibration sources with similar precision (e.g.\ a secondary LFC or an absorption cell). 
    
    \item An unexpected digitisation issue introduced weak flux artefacts into ESPRESSO spectra of faint targets (such as quasars) taken before 19 May 2022. Repeating some faint target observations may be useful in understanding the effect so it can be efficiently and automatically removed from the affected datasets. Currently, that data must be cleaned manually (Sec.~\ref{sec:feb}).

    \item The experience from using HARPS and ESPRESSO for high precision spectroscopic measurements highlights the importance of understanding the full data production and data analysis chains. Further work is required before these instruments can be used optimally. 

    \item When fitting multiple species simultaneously to measure $\daoaline$, only compound broadening and SpIC (or AICc) should be used for model building \citep{Webb2022}. Turbulent broadening should not be used for simultaneous fits to multiple species
    as doing so will likely bias the measurement \citep{Webb2022}. 

    \item Uncertainty on $\daoaline$ due to model non-uniqueness is sometimes comparable to or even larger than the statistical uncertainty \citep{Lee2021aivpfit,Lee2021nonunique,Webb2021,Webb2022}. For any particular absorption system, multiple models should be produced to quantify this uncertainty \citep{Webb2022}. Interestingly, using compound broadening and SpIC can eliminate non-uniqueness in some observations \citep{Lee2021nonunique}. A future analysis using archival (e.g.\ UVES) and novel ESPRESSO observations may be used to explore the role of spectral resolution and S/N.

    \item Non-uniqueness may be endemic in quasar absorption systems. Conclusions on the variability of fundamental constants will then have to be made on a large statistical sample instead of on individual quasar absorption systems \citep{Webb2022}.

\end{enumerate}

\begin{acknowledgements}
DM is also supported by the INFN PD51 INDARK grant. 
\end{acknowledgements}

\bibliographystyle{aa}
\bibliography{main}

\begin{thebibliography}{51}
\expandafter\ifx\csname natexlab\endcsname\relax\def\natexlab#1{#1}\fi

\bibitem[{{Akaike}(1974)}]{Akaike1974}
{Akaike}, H. 1974, IEEE Transactions on Automatic Control, 19, 716

\bibitem[{{Baade} {et~al.}(2009){Baade}, {Balestra}, {Cumani}, {Eschbaumer},
  {Finger}, {Geimer}, {Mehrgan}, {Meyer}, {Stegmeier}, {Reyes}, \&
  {Todorovic}}]{Baade2009}
{Baade}, D., {Balestra}, A., {Cumani}, C., {et~al.} 2009, The Messenger, 136,
  20

\bibitem[{{Bainbridge} \& {Webb}(2017)}]{Bainbridge2017}
{Bainbridge}, M.~B. \& {Webb}, J.~K. 2017, \mnras, 468, 1639

\bibitem[{{Berengut} {et~al.}(2013){Berengut}, {Flambaum}, {Ong}, {Webb},
  {Barrow}, {Barstow}, {Preval}, \& {Holberg}}]{Berengut2013}
{Berengut}, J.~C., {Flambaum}, V.~V., {Ong}, A., {et~al.} 2013, \prl, 111,
  010801

\bibitem[{{Bouchy} {et~al.}(2017){Bouchy}, {Doyon}, {Artigau}, {Melo},
  {Hernandez}, {Wildi}, {Delfosse}, {Lovis}, {Figueira}, {Canto Martins},
  {Gonz{\'a}lez Hern{\'a}ndez}, {Thibault}, {Reshetov}, {Pepe}, {Santos}, {de
  Medeiros}, {Rebolo}, {Abreu}, {Adibekyan}, {Bandy}, {Benz}, {Blind},
  {Bohlender}, {Boisse}, {Bovay}, {Broeg}, {Brousseau}, {Cabral}, {Chazelas},
  {Cloutier}, {Coelho}, {Conod}, {Cumming}, {Delabre}, {Genolet}, {Hagelberg},
  {Jayawardhana}, {K{\"a}ufl}, {Lafreni{\`e}re}, {de Castro Le{\~a}o}, {Malo},
  {de Medeiros Martins}, {Matthews}, {Metchev}, {Oshagh}, {Ouellet}, {Parro},
  {Rasilla Pi{\~n}eiro}, {Santos}, {Sarajlic}, {Segovia}, {Sordet}, {Udry},
  {Valencia}, {Vall{\'e}e}, {Venn}, {Wade}, \& {Saddlemyer}}]{Bouchy2017}
{Bouchy}, F., {Doyon}, R., {Artigau}, {\'E}., {et~al.} 2017, The Messenger,
  169, 21

\bibitem[{Bozdogan(1987)}]{Bozdogan1987}
Bozdogan, H. 1987, Psychometrika, 52, 345

\bibitem[{{Carswell} \& {Webb}(2014)}]{Carswell2014}
{Carswell}, R.~F. \& {Webb}, J.~K. 2014, {VPFIT: Voigt profile fitting
  program}, Astrophysics Source Code Library, record ascl:1408.015

\bibitem[{{Dumont} \& {Webb}(2017)}]{Dumont2017}
{Dumont}, V. \& {Webb}, J.~K. 2017, \mnras, 468, 1568

\bibitem[{{Dzuba} {et~al.}(1999){Dzuba}, {Flambaum}, \& {Webb}}]{Dzuba1999}
{Dzuba}, V.~A., {Flambaum}, V.~V., \& {Webb}, J.~K. 1999, \prl, 82, 888

\bibitem[{{Evans} {et~al.}(2014){Evans}, {Murphy}, {Whitmore}, {Misawa},
  {Centurion}, {D'Odorico}, {Lopez}, {Martins}, {Molaro}, {Petitjean},
  {Rahmani}, {Srianand}, \& {Wendt}}]{Evans2014}
{Evans}, T.~M., {Murphy}, M.~T., {Whitmore}, J.~B., {et~al.} 2014, \mnras, 445,
  128

\bibitem[{{H{\"a}nsch}(2006)}]{Haensch2006}
{H{\"a}nsch}, T.~W. 2006, Reviews of Modern Physics, 78, 1297

\bibitem[{{Hees} {et~al.}(2020){Hees}, {Do}, {Roberts}, {Ghez}, {Nishiyama},
  {Bentley}, {Gautam}, {Jia}, {Kara}, {Lu}, {Saida}, {Sakai}, {Takahashi}, \&
  {Takamori}}]{Hees2020}
{Hees}, A., {Do}, T., {Roberts}, B.~M., {et~al.} 2020, \prl, 124, 081101

\bibitem[{{Hu} {et~al.}(2021){Hu}, {Webb}, {Ayres}, {Bainbridge}, {Barrow},
  {Barstow}, {Berengut}, {Carswell}, {Dumont}, {Dzuba}, {Flambaum}, {Lee},
  {Reindl}, {Preval}, \& {Tchang-Brillet}}]{Hu2021}
{Hu}, J., {Webb}, J.~K., {Ayres}, T.~R., {et~al.} 2021, \mnras, 500, 1466

\bibitem[{Hurvich \& Tsai(1989)}]{Hurvich1989}
Hurvich, C.~M. \& Tsai, C.-L. 1989, Biometrika, 76, 297

\bibitem[{{King} {et~al.}(2012){King}, {Webb}, {Murphy}, {Flambaum},
  {Carswell}, {Bainbridge}, {Wilczynska}, \& {Koch}}]{King2012}
{King}, J.~A., {Webb}, J.~K., {Murphy}, M.~T., {et~al.} 2012, \mnras, 422, 3370

\bibitem[{{Kotu{\v{s}}} {et~al.}(2017){Kotu{\v{s}}}, {Murphy}, \&
  {Carswell}}]{Kotus2017}
{Kotu{\v{s}}}, S.~M., {Murphy}, M.~T., \& {Carswell}, R.~F. 2017, \mnras, 464,
  3679

\bibitem[{{Lee} {et~al.}(2021{\natexlab{a}}){Lee}, {Webb}, {Carswell}, \&
  {Milakovi{\'c}}}]{Lee2021aivpfit}
{Lee}, C.-C., {Webb}, J.~K., {Carswell}, R.~F., \& {Milakovi{\'c}}, D.
  2021{\natexlab{a}}, \mnras, 504, 1787

\bibitem[{{Lee} {et~al.}(2021{\natexlab{b}}){Lee}, {Webb}, {Milakovi{\'c}}, \&
  {Carswell}}]{Lee2021nonunique}
{Lee}, C.-C., {Webb}, J.~K., {Milakovi{\'c}}, D., \& {Carswell}, R.~F.
  2021{\natexlab{b}}, \mnras, 507, 27

\bibitem[{{Liddle}(2007)}]{Liddle2007}
{Liddle}, A.~R. 2007, \mnras, 377, L74

\bibitem[{{Marconi} {et~al.}(2022){Marconi}, {Abreu}, {Adibekyan}, {Alberti},
  {Albrecht}, {Alcaniz}, {Aliverti}, {Allende Prieto}, {Alvarado G{\'o}mez},
  {Amado}, {Amate}, {Andersen}, {Artigau}, {Baker}, {Baldini}, {Balestra},
  {Barnes}, {Baron}, {Barros}, {Bauer}, {Beaulieu}, {Bellido-Tirado},
  {Benneke}, {Bensby}, {Bergin}, {Biazzo}, {Bik}, {Birkby}, {Blind}, {Boisse},
  {Bolmont}, {Bonaglia}, {Bonfils}, {Borsa}, {Brandeker}, {Brandner}, {Broeg},
  {Brogi}, {Brousseau}, {Brucalassi}, {Brynnel}, {Buchhave}, {Buscher},
  {Cabral}, {Calderone}, {Calvo-Ortega}, {Canto Martins}, {Cantalloube},
  {Carbonaro}, {Chauvin}, {Chazelas}, {Cheffot}, {Cheng}, {Chiavassa},
  {Christensen}, {Cirami}, {Cook}, {Cooke}, {Coretti}, {Covino}, {Cowan},
  {Cresci}, {Cristiani}, {Cunha Parro}, {Cupani}, {D'Odorico}, {de Castro
  Le{\~a}o}, {De Cia}, {De Medeiros}, {Debras}, {Debus}, {Demangeon},
  {Dessauges-Zavadsky}, {Di Marcantonio}, {Dionies}, {Doyon}, {Dunn},
  {Ehrenreich}, {Faria}, {Feruglio}, {Fisher}, {Fontana}, {Fumagalli}, {Fusco},
  {Fynbo}, {Gabella}, {Gaessler}, {Gallo}, {Gao}, {Genolet}, {Genoni},
  {Giacobbe}, {Giro}, {Gon{\c{c}}alves}, {Gonzalez}, {Gonz{\'a}lez
  Hern{\'a}ndez}, {Gracia T{\'e}mich}, {Haehnelt}, {Haniff}, {Hatzes},
  {Helled}, {Hoeijmakers}, {Huke}, {J{\"a}rvinen}, {J{\"a}rvinen}, {Kaminski},
  {Korn}, {Kouach}, {Kowzan}, {Kreidberg}, {Landoni}, {Lanotte}, {Lavail},
  {Li}, {Liske}, {Lovis}, {Lucatello}, {Lunney}, {MacIntosh}, {Madhusudhan},
  {Magrini}, {Maiolino}, {Malo}, {Man}, {Marquart}, {Marques}, {Martins},
  {Martins}, {Maslowski}, {Mason}, {Mason}, {McCracken}, {Mergo}, {Micela},
  {Mitchell}, {Molli{\`e}re}, {Monteiro}, {Montgomery}, {Mordasini}, {Morin},
  {Mucciarelli}, {Murphy}, {N'Diaye}, {Neichel}, {Niedzielski}, {Niemczura},
  {Nortmann}, {Noterdaeme}, {Nunes}, {Oggioni}, {Oliva}, {{\"O}nel}, {Origlia},
  {{\"O}stlin}, {Palle}, {Papaderos}, {Pariani}, {Pe{\~n}ate Castro}, {Pepe},
  {Perreault Levasseur}, {Petit}, {Pino}, {Piqueras}, {Pollo}, {Poppenhaeger},
  {Quirrenbach}, {Rauscher}, {Rebolo}, {Redaelli}, {Reffert}, {Reid},
  {Reiners}, {Richter}, {Riva}, {Rivoire}, {Rodr{\'\i}guez-L{\'o}pez},
  {Roederer}, {Romano}, {Rousseau}, {Rowe}, {Salvadori}, {Santos}, {Santos
  Diaz}, {Sanz-Forcada}, {Sarajlic}, {Sauvage}, {Sch{\"a}fer}, {Schiavon},
  {Schmidt}, {Selmi}, {Sivanandam}, {Sordet}, {Sordo}, {Sortino}, {Sosnowska},
  {Sousa}, {Stempels}, {Strassmeier}, {Su{\'a}rez Mascare{\~n}o}, {Sulich},
  {Sun}, {Tanvir}, {Tenegi-Sangin{\'e}s}, {Thibault}, {Thompson}, {Tozzi},
  {Turbet}, {Vall{\'e}e}, {Varas}, {Venn}, {V{\'e}ran}, {Verma}, {Viel},
  {Wade}, {Waring}, {Weber}, {Weder}, {Wehbe}, {Weingrill}, {Woche}, {Xompero},
  {Zackrisson}, {Zanutta}, {Zapatero Osorio}, {Zechmeister}, \&
  {Zimara}}]{Marconi2022}
{Marconi}, A., {Abreu}, M., {Adibekyan}, V., {et~al.} 2022, in Society of
  Photo-Optical Instrumentation Engineers (SPIE) Conference Series, Vol. 12184,
  Ground-based and Airborne Instrumentation for Astronomy IX, ed. C.~J.
  {Evans}, J.~J. {Bryant}, \& K.~{Motohara}, 1218424

\bibitem[{{Martins}(2017)}]{Martins2017}
{Martins}, C.~J.~A.~P. 2017, Reports on Progress in Physics, 80, 126902

\bibitem[{{Mayor} {et~al.}(2003){Mayor}, {Pepe}, {Queloz}, {Bouchy},
  {Rupprecht}, {Lo Curto}, {Avila}, {Benz}, {Bertaux}, {Bonfils}, {Dall},
  {Dekker}, {Delabre}, {Eckert}, {Fleury}, {Gilliotte}, {Gojak}, {Guzman},
  {Kohler}, {Lizon}, {Longinotti}, {Lovis}, {Megevand}, {Pasquini}, {Reyes},
  {Sivan}, {Sosnowska}, {Soto}, {Udry}, {van Kesteren}, {Weber}, \&
  {Weilenmann}}]{Mayor2003}
{Mayor}, M., {Pepe}, F., {Queloz}, D., {et~al.} 2003, The Messenger, 114, 20

\bibitem[{{Milakovi{\'c}} {et~al.}(2021){Milakovi{\'c}}, {Lee}, {Carswell},
  {Webb}, {Molaro}, \& {Pasquini}}]{Milakovic2021}
{Milakovi{\'c}}, D., {Lee}, C.-C., {Carswell}, R.~F., {et~al.} 2021, \mnras,
  500, 1

\bibitem[{{Milakovi{\'c}} {et~al.}(2020){Milakovi{\'c}}, {Pasquini}, {Webb}, \&
  {Lo Curto}}]{Milakovic2020}
{Milakovi{\'c}}, D., {Pasquini}, L., {Webb}, J.~K., \& {Lo Curto}, G. 2020,
  \mnras, 493, 3997

\bibitem[{{Murphy}(2018)}]{Murphy2018UVESpopler}
{Murphy}, M. 2018, {Mtmurphy77/Uves\_Popler: Uves\_Popler: Post-Pipeline
  Echelle Reduction Software}, Zenodo

\bibitem[{{Murphy} {et~al.}(2022{\natexlab{a}}){Murphy}, {Berke}, {Liu},
  {Flynn}, {Lehmann}, {Dzuba}, \& {Flambaum}}]{Murphy2022_solartwins}
{Murphy}, M.~T., {Berke}, D.~A., {Liu}, F., {et~al.} 2022{\natexlab{a}}, arXiv
  e-prints, arXiv:2211.05150

\bibitem[{{Murphy} \& {Cooksey}(2017)}]{Murphy2017}
{Murphy}, M.~T. \& {Cooksey}, K.~L. 2017, \mnras, 471, 4930

\bibitem[{{Murphy} {et~al.}(2019){Murphy}, {Kacprzak}, {Savorgnan}, \&
  {Carswell}}]{Murphy2019}
{Murphy}, M.~T., {Kacprzak}, G.~G., {Savorgnan}, G. A.~D., \& {Carswell}, R.~F.
  2019, \mnras, 482, 3458

\bibitem[{{Murphy} {et~al.}(2022{\natexlab{b}}){Murphy}, {Molaro}, {Leite},
  {Cupani}, {Cristiani}, {D'Odorico}, {G{\'e}nova Santos}, {Martins},
  {Milakovi{\'c}}, {Nunes}, {Schmidt}, {Pepe}, {Rebolo}, {Santos}, {Sousa},
  {Zapatero Osorio}, {Amate}, {Adibekyan}, {Alibert}, {Prieto}, {Baldini},
  {Benz}, {Bouchy}, {Cabral}, {Dekker}, {Di Marcantonio}, {Ehrenreich},
  {Figueira}, {Gonz{\'a}lez Hern{\'a}ndez}, {Landoni}, {Lovis}, {Lo Curto},
  {Manescau}, {M{\'e}gevand}, {Mehner}, {Micela}, {Pasquini}, {Poretti},
  {Riva}, {Sozzetti}, {Mascare{\~n}o}, {Udry}, \& {Zerbi}}]{Murphy2022}
{Murphy}, M.~T., {Molaro}, P., {Leite}, A. C.~O., {et~al.} 2022{\natexlab{b}},
  \aap, 658, A123

\bibitem[{{Murphy} {et~al.}(2022{\natexlab{c}}){Murphy}, {Molaro}, {Schmidt},
  {Martins}, {da Fonseca}, {Milakovi{\'c}}, {Cupani}, {Cristiani}, {D'Odorico},
  {Barreiro}, {G{\'e}nova Santos}, {Leite}, {Marques}, {Nunes}, {Pepe},
  {Rebolo}, {Santos}, {Sousa}, {Zapatero Osorio}, {Amate}, {Adibekyan},
  {Alibert}, {Allende Prieto}, {Baldini}, {Benz}, {Bouchy}, {Cabral}, {Dekker},
  {Di Marcantonio}, {Ehrenreich}, {Figueira}, {Gonz{\'a}lez Hern{\'a}ndez},
  {Landoni}, {Lovis}, {Lo Curto}, {Manescau}, {M{\'e}gevand}, {Mehner},
  {Micela}, {Pasquini}, {Poretti}, {Riva}, {Sozzetti}, {Su{\'a}rez
  Mascare{\~n}o}, {Udry}, \& {Zerbi}}]{Murphy2022_messenger}
{Murphy}, M.~T., {Molaro}, P., {Schmidt}, T.~M., {et~al.} 2022{\natexlab{c}},
  The Messenger, 188, 15

\bibitem[{{Obrzud} {et~al.}(2019){Obrzud}, {Brasch}, {Voumard}, {Stroganov},
  {Geiselmann}, {Wildi}, {Pepe}, {Lecomte}, \& {Herr}}]{Obrzud2019}
{Obrzud}, E., {Brasch}, V., {Voumard}, T., {et~al.} 2019, Optics Letters, 44,
  5290

\bibitem[{{Obrzud} {et~al.}(2021){Obrzud}, {Denis}, {Sattari}, {Choong},
  {Kundermann}, {Dubochet}, {Despont}, {Lecomte}, {Ghadimi}, \&
  {Brasch}}]{Obrzud2021}
{Obrzud}, E., {Denis}, S., {Sattari}, H., {et~al.} 2021, APL Photonics, 6,
  121303

\bibitem[{{Pepe} {et~al.}(2021){Pepe}, {Cristiani}, {Rebolo}, {Santos},
  {Dekker}, {Cabral}, {Di Marcantonio}, {Figueira}, {Lo Curto}, {Lovis},
  {Mayor}, {M{\'e}gevand}, {Molaro}, {Riva}, {Zapatero Osorio}, {Amate},
  {Manescau}, {Pasquini}, {Zerbi}, {Adibekyan}, {Abreu}, {Affolter}, {Alibert},
  {Aliverti}, {Allart}, {Allende Prieto}, {{\'A}lvarez}, {Alves}, {Avila},
  {Baldini}, {Bandy}, {Barros}, {Benz}, {Bianco}, {Borsa}, {Bourrier},
  {Bouchy}, {Broeg}, {Calderone}, {Cirami}, {Coelho}, {Conconi}, {Coretti},
  {Cumani}, {Cupani}, {D'Odorico}, {Damasso}, {Deiries}, {Delabre},
  {Demangeon}, {Dumusque}, {Ehrenreich}, {Faria}, {Fragoso}, {Genolet},
  {Genoni}, {G{\'e}nova Santos}, {Gonz{\'a}lez Hern{\'a}ndez}, {Hughes},
  {Iwert}, {Kerber}, {Knudstrup}, {Landoni}, {Lavie}, {Lillo-Box}, {Lizon},
  {Maire}, {Martins}, {Mehner}, {Micela}, {Modigliani}, {Monteiro}, {Monteiro},
  {Moschetti}, {Murphy}, {Nunes}, {Oggioni}, {Oliveira}, {Oshagh}, {Pall{\'e}},
  {Pariani}, {Poretti}, {Rasilla}, {Rebord{\~a}o}, {Redaelli}, {Santana
  Tschudi}, {Santin}, {Santos}, {S{\'e}gransan}, {Schmidt}, {Segovia},
  {Sosnowska}, {Sozzetti}, {Sousa}, {Span{\`o}}, {Su{\'a}rez Mascare{\~n}o},
  {Tabernero}, {Tenegi}, {Udry}, \& {Zanutta}}]{Pepe2021}
{Pepe}, F., {Cristiani}, S., {Rebolo}, R., {et~al.} 2021, \aap, 645, A96

\bibitem[{{Probst} {et~al.}(2014){Probst}, {Lo Curto}, {Avila}, {Canto
  Martins}, {de Medeiros}, {Esposito}, {Gonz{\'a}lez Hern{\'a}ndez},
  {H{\"a}nsch}, {Holzwarth}, {Kerber}, {Le{\~a}o}, {Manescau}, {Pasquini},
  {Rebolo-L{\'o}pez}, {Steinmetz}, {Udem}, \& {Wu}}]{Probst2014}
{Probst}, R.~A., {Lo Curto}, G., {Avila}, G., {et~al.} 2014, in Society of
  Photo-Optical Instrumentation Engineers (SPIE) Conference Series, Vol. 9147,
  Ground-based and Airborne Instrumentation for Astronomy V, ed. S.~K.
  {Ramsay}, I.~S. {McLean}, \& H.~{Takami}, 91471C

\bibitem[{{Probst} {et~al.}(2020){Probst}, {Milakovi{\'c}},
  {Toledo-Padr{\'o}n}, {Lo Curto}, {Avila}, {Brucalassi}, {Canto Martins}, {de
  Castro Le{\~a}o}, {Esposito}, {Gonz{\'a}lez Hern{\'a}ndez}, {Grupp},
  {H{\"a}nsch}, {Kellermann}, {Kerber}, {Mandel}, {Manescau}, {Pozna},
  {Rebolo}, {de Medeiros}, {Steinmetz}, {Su{\'a}rez Mascare{\~n}o}, {Udem},
  {Urrutia}, {Wu}, {Pasquini}, \& {Holzwarth}}]{Probst2020}
{Probst}, R.~A., {Milakovi{\'c}}, D., {Toledo-Padr{\'o}n}, B., {et~al.} 2020,
  Nature Astronomy, 4, 603

\bibitem[{{Rahmani} {et~al.}(2013){Rahmani}, {Wendt}, {Srianand}, {Noterdaeme},
  {Petitjean}, {Molaro}, {Whitmore}, {Murphy}, {Centurion}, {Fathivavsari},
  {D'Odorico}, {Evans}, {Levshakov}, {Lopez}, {Martins}, {Reimers}, \&
  {Vladilo}}]{Rahmani2013}
{Rahmani}, H., {Wendt}, M., {Srianand}, R., {et~al.} 2013, \mnras, 435, 861

\bibitem[{{Rasmussen} \& {Williams}(2006)}]{Rasmussen2006}
{Rasmussen}, C.~E. \& {Williams}, C. K.~I. 2006, {Gaussian Processes for
  Machine Learning}

\bibitem[{{Schmidt} {et~al.}(2021{\natexlab{a}}){Schmidt}, {Bouchy}, {Brasch},
  {Herr}, {Pepe}, \& {Lovis}}]{Schmidt2021bluves}
{Schmidt}, T., {Bouchy}, F., {Brasch}, V., {et~al.} 2021{\natexlab{a}}, in
  European Planetary Science Congress, EPSC2021--323

\bibitem[{{Schmidt} {et~al.}(2021{\natexlab{b}}){Schmidt}, {Molaro}, {Murphy},
  {Lovis}, {Cupani}, {Cristiani}, {Pepe}, {Rebolo}, {Santos}, {Abreu},
  {Adibekyan}, {Alibert}, {Aliverti}, {Allart}, {Allende Prieto}, {Alves},
  {Baldini}, {Broeg}, {Cabral}, {Calderone}, {Cirami}, {Coelho}, {Coretti},
  {D'Odorico}, {Di Marcantonio}, {Ehrenreich}, {Figueira}, {Genoni},
  {G{\'e}nova Santos}, {Gonz{\'a}lez Hern{\'a}ndez}, {Kerber}, {Landoni},
  {Leite}, {Lizon}, {Lo Curto}, {Manescau}, {Martins}, {Meg{\'e}vand},
  {Mehner}, {Micela}, {Modigliani}, {Monteiro}, {Monteiro}, {Mueller}, {Nunes},
  {Oggioni}, {Oliveira}, {Pariani}, {Pasquini}, {Redaelli}, {Riva}, {Santos},
  {Sosnowska}, {Sousa}, {Sozzetti}, {Su{\'a}rez Mascare{\~n}o}, {Udry},
  {Zapatero Osorio}, \& {Zerbi}}]{Schmidt2021}
{Schmidt}, T.~M., {Molaro}, P., {Murphy}, M.~T., {et~al.} 2021{\natexlab{b}},
  \aap, 646, A144

\bibitem[{{Schwarz}(1978)}]{Schwarz1978}
{Schwarz}, G. 1978, Annals of Statistics, 6, 461

\bibitem[{{Udem} {et~al.}(2002){Udem}, {Holzwarth}, \& {H{\"a}nsch}}]{Udem2002}
{Udem}, T., {Holzwarth}, R., \& {H{\"a}nsch}, T.~W. 2002, \nat, 416, 233

\bibitem[{{Uzan}(2011)}]{Uzan2011}
{Uzan}, J.-P. 2011, Living Reviews in Relativity, 14, 2

\bibitem[{{Webb} {et~al.}(1999){Webb}, {Flambaum}, {Churchill}, {Drinkwater},
  \& {Barrow}}]{Webb1999}
{Webb}, J.~K., {Flambaum}, V.~V., {Churchill}, C.~W., {Drinkwater}, M.~J., \&
  {Barrow}, J.~D. 1999, \prl, 82, 884

\bibitem[{{Webb} {et~al.}(2021){Webb}, {Lee}, {Carswell}, \&
  {Milakovi{\'c}}}]{Webb2021}
{Webb}, J.~K., {Lee}, C.-C., {Carswell}, R.~F., \& {Milakovi{\'c}}, D. 2021,
  \mnras, 501, 2268

\bibitem[{{Webb} {et~al.}(2022){Webb}, {Lee}, \& {Milakovi{\'c}}}]{Webb2022}
{Webb}, J.~K., {Lee}, C.-C., \& {Milakovi{\'c}}, D. 2022, Universe, 8, 266

\bibitem[{{Webb} {et~al.}(2001){Webb}, {Murphy}, {Flambaum}, {Dzuba}, {Barrow},
  {Churchill}, {Prochaska}, \& {Wolfe}}]{Webb2001}
{Webb}, J.~K., {Murphy}, M.~T., {Flambaum}, V.~V., {et~al.} 2001, \prl, 87,
  091301

\bibitem[{{Whitmore} \& {Murphy}(2015)}]{Whitmore2015}
{Whitmore}, J.~B. \& {Murphy}, M.~T. 2015, \mnras, 447, 446

\bibitem[{{Wilczynska} {et~al.}(2015){Wilczynska}, {Webb}, {King}, {Murphy},
  {Bainbridge}, \& {Flambaum}}]{Wilczynska2015}
{Wilczynska}, M.~R., {Webb}, J.~K., {King}, J.~A., {et~al.} 2015, \mnras, 454,
  3082

\bibitem[{{Wilken} {et~al.}(2012){Wilken}, {Curto}, {Probst}, {Steinmetz},
  {Manescau}, {Pasquini}, {Gonz{\'a}lez Hern{\'a}ndez}, {Rebolo}, {H{\"a}nsch},
  {Udem}, \& {Holzwarth}}]{Wilken2012}
{Wilken}, T., {Curto}, G.~L., {Probst}, R.~A., {et~al.} 2012, \nat, 485, 611

\bibitem[{{Wilken} {et~al.}(2010){Wilken}, {Lovis}, {Manescau}, {Steinmetz},
  {Pasquini}, {Lo Curto}, {H{\"a}nsch}, {Holzwarth}, \& {Udem}}]{Wilken2010}
{Wilken}, T., {Lovis}, C., {Manescau}, A., {et~al.} 2010, \mnras, 405, L16

\bibitem[{{Zhao} {et~al.}(2021){Zhao}, {Lo Curto}, {Pasquini}, {Gonz{\'a}lez
  Hern{\'a}ndez}, {De Medeiros}, {Canto Martins}, {Le{\~a}o}, {Rebolo},
  {Su{\'a}rez Mascare{\~n}o}, {Esposito}, {Manescau}, {Steinmetz}, {Udem},
  {Probst}, {Holzwarth}, \& {Zhao}}]{Zhao2021}
{Zhao}, F., {Lo Curto}, G., {Pasquini}, L., {et~al.} 2021, \aap, 645, A23

\end{thebibliography}

\end{document}